\documentclass[11pt]{article}
\usepackage{amssymb,amsmath,amsthm,amscd,latexsym}
\usepackage{mathrsfs}
\usepackage{mathrsfs}
\usepackage{amsfonts}
\usepackage{amsmath}
\usepackage{amssymb}
\usepackage{amscd}
\usepackage{enumitem}
\usepackage[all]{xy}
\usepackage{CJK}
\usepackage{xypic}
\usepackage{graphicx}
\usepackage{epstopdf}

\renewcommand{\paragraph}{\roman{paragraph}}
\input cyracc.def

\newcommand{\Z}{\mathbb{Z}}
\newcommand{\F}{\mathbb{F}}

\begin{document}
\begin{CJK*}{GBK}{song}\CJKtilde
\title{\bf Skew Cyclic codes over $\F_q+u\F_q+v\F_q+uv\F_q$
\thanks{Corresponding author: Minjia Shi. \textbf{The original manuscript was first submitted for reviewing on 4nd December 2014.}
This research is supported by NNSF of China (61202068),
Talented youth Fund of Anhui Province Universities (2012SQRL020ZD).}}
\author{\small{Ting YAO, Minjia SHI}\\ \small{School of Mathematical Sciences of Anhui University, China} \and \small{Patrick SOL\'E}\\ \small{Telecom Paris Tech, France and King Abdulaziz University, Saudi Arabia}}
\date{}
\maketitle
\end{CJK*}

\begin{CJK}{GBK}{song}
{\bf Abstract:} {\normalsize   In this paper, we study skew cyclic codes over the ring $R=\F_q+u\F_q+v\F_q+uv\F_q$, where $u^{2}=u,v^{2}=v,uv=vu$, $q=p^{m}$ and $p$ is an odd prime. We investigate the structural properties of skew cyclic codes over $R$ through a decomposition theorem. Furthermore, we give a formula for the number of skew cyclic codes of length $n$ over $R.$}


\section{Introduction}

\hspace*{0.6cm}Cyclic codes form an important subclass of linear block codes, studied from the fifties onward.
Their clear algebraic structures as ideals of a quotient ring of a polynomial ring makes for an easy encoding. A landmark paper [11] has shown that some important binary nonlinear codes with excellent error-correcting capabilities can be identified as images of linear codes over $\Z_{4}$ under the Gray map.

Recently, in [3], D. Boucher et al. gave skew cyclic codes defined by using the skew polynomial ring with an automorphism $\theta$ over the finite field with $q$ elements. The definition generalizes the concept of cyclic codes over non-commutative polynomial rings. Soon afterwards, D. Boucher et al. studied skew constacyclic codes in [5]. Later, in [4], some important results on the duals of the skew cyclic codes over $\F_{q}[x; \theta]$ are given. In [12], I. Siap et al. presented the structure of skew cyclic codes of arbitrary length. Further, S. Jitman et al. in [10] defined skew constacyclic codes over the skew polynomial ring with coefficients from finite rings. In [1], T. Abualrub and P. Seneviratne studied skew cyclic codes over ring $\F_{2}+v\F_{2}$ with $v^{2}=v$. Moreover, J. Gao [6] and F. Gursoy et al. [8] presented skew cyclic codes over $\F_{p}+v\F_{p}$ and $\F_{q}+v\F_{q}$ with different automorphisms, respectively. In [7], J. Gao et al. also studied skew generalized quasi-cyclic codes over finite fields.

In this article, we mainly study skew cyclic codes over ring $R=\F_q+u\F_q+v\F_q+uv\F_q$, where $u^{2}=u,v^{2}=v,uv=vu$ and $q=p^{m}.$

In our work, the automorphism $\theta$ on the ring $R$ is defined to be $$\theta(b_{0}+b_{1}u+b_{2}v+b_{3}uv)=b_{0}^{p}+b_{2}^{p}u+b_{1}^{p}v+b_{3}^{p}uv,$$
for all $b_{0}+b_{1}u+b_{2}v+b_{3}uv\in R$, where $b_{i}\in\F_q,$ and $i=0,1,2,3.$ In fact, for any $a_{1}\eta_{1}+a_{2}\eta_{2}+a_{3}\eta_{3}+a_{4}\eta_{4}\in R$, we have
  $$\theta(a_{1}\eta_{1}+a_{2}\eta_{2}+a_{3}\eta_{3}+a_{4}\eta_{4})=
  \theta(a_{1})\eta_{1}+\theta(a_{2})\eta_{2}+\theta(a_{4})\eta_{3}+\theta(a_{3})\eta_{4}.$$
Note that if $m$ is even, the order of the ring automorphism $|\langle\theta\rangle|$ is $m$, otherwise, $2m$.

The material is organized as follows. In Section 2, we show the basics of codes over ring $R$ that we need for further reference. Section 3 derives the structure of linear codes over $R$. In Section 4, we introduce skew cyclic codes over ring $R$ and give the structural properties of skew cyclic codes over $R$ through a decomposition theorem. Section 5, we give a example to illustrate the discussed results.
\section{Preliminary}

\hspace*{0.6cm}Let $\F_q$ be a finite field with $q$ elements, where $q=p^{m}$, $p$ is an odd prime.
Throughout, we let $R$ denote the commutative ring $\F_q+u\F_q+v\F_q+uv\F_q,$ where $u^{2}=u, v^{2}=v,$ and $uv=vu.$ Let $\eta_{1}=1-u-v+uv,$ $\eta_{2}=uv,$ $\eta_{3}=u-uv,$ $\eta_{4}=v-uv.$ It is easy to verify that $\eta_{i}^{2}=\eta_{i}, \eta_{i}\eta_{j}=0,$ and $\sum_{k=1}^{4}\eta_{k}=1,$ where $i, j=1, 2, 3, 4,$ and $i\neq j$. According to [2], we have $R=\eta_{1}R\oplus\eta_{2}R\oplus\eta_{3}R\oplus\eta_{4}R$. By calculating, we can easily obtain that  $\eta_{i}R\cong\F_q$, $i=1,2,3,4.$ Therefore, for any $r\in R$, $r$ can be expressed uniquely as $r=\sum_{i=1}^{4}\eta_{i}a_{i}$, where $a_{i}\in \F_{q}$ for $i=1, 2, 3, 4.$\

We recall the definition of the Gray map over $R$ in [13]
\begin{eqnarray*}
 \Phi: R=\F_q+u\F_q+v\F_q+uv\F_q&\rightarrow&\F_q^{4}\\
\eta_{1}a+\eta_{2}b+\eta_{3}c+\eta_{4}d&\rightarrow&(a,a+b,a+c,a+b+c+d).
\end{eqnarray*}
Equivalently, if $r=a'+b'u+c'v+d'uv\in R$, then $$\Phi(r)=(a', 2a'+b'+c'+d', 2a'+b', 4a'+2b'+2c'+d').$$
This map can be naturally extended to the case over $R^{n}$.

For any element $r=a+bu+cv+duv\in R,$ we define the Lee weight of $r$ as
  $w_{L}(r)=w_{H}(a,a+b,a+c,a+b+c+d)$, where $w_{H}$ denotes the ordinary Hamming weight for $q$-ary codes. The Lee distance of $r\in R$ can be similarly defined.

From the definition of the Gray map $\Phi$, we can easily check that $\Phi$ is $\F_{q}$-linear and it is also a distance-reserving isometry from $(R^{n},d_{L})$ to $(F^{4n}_{q}
,d_{H}),$
where $d_{L}$ and $d_{H}$ denote the Lee and Hamming distance in $R^{n}$ and $F^{4n}_{q}$, respectively.

\section{Linear codes over $R$ }

\hspace*{0.6cm}In this section, we mainly show some familiar structural properties of $R$. The proofs of the following theorems can be found in [13], so we omit them here.

If $A_{i}~(i=1, 2, 3, 4)$ are codes over $R$, we denote their direct sum by
 $$A_{1}\oplus A_{2}\oplus A_{3}\oplus A_{4}=\{a_{1}+a_{2}+a_{3}+a_{4}|a_{i}\in A_{i},i=1,2,3,4\}.$$

\textbf{Definition 3.1} Let $C$ be a linear code of length $n$ over $R$, we define that
$$C_{1}=\{\textbf{a}\in \F_q^{n}|\exists \textbf{b,c,d}\in\F_q^{n}|\eta_{1}
\textbf{a}+\eta_{2}\textbf{b}+\eta_{3}\textbf{c}+\eta_{4}\textbf{d}\in C\},$$
$$C_{2}=\{\textbf{b}\in \F_q^{n}|\exists \textbf{a,c,d}\in\F_q^{n}|\eta_{1}
\textbf{a}+\eta_{2}\textbf{b}+\eta_{3}\textbf{c}+\eta_{4}\textbf{d}\in C\},$$
$$C_{3}=\{\textbf{c}\in \F_q^{n}|\exists \textbf{a,b,d}\in\F_q^{n}|\eta_{1}
\textbf{a}+\eta_{2}\textbf{b}+\eta_{3}\textbf{c}+\eta_{4}\textbf{d}\in C\},$$
$$C_{4}=\{\textbf{d}\in \F_q^{n}|\exists \textbf{a,b,c}\in\F_q^{n}|\eta_{1}
\textbf{a}+\eta_{2}\textbf{b}+\eta_{3}\textbf{c}+\eta_{4}\textbf{d}\in C\}.$$
It is clear that $C_{i}~(i=1, 2, 3, 4)$ are linear codes over $\F_q^{n}$. Furthermore,
$C=\eta_{1}C_{1}\oplus\eta_{2}C_{2}\oplus\eta_{3}C_{3}\oplus\eta_{4}C_{4},$ and $|C|=|C_{1}|\cdot|C_{2}|\cdot|C_{3}|\cdot|C_{4}|.$ Throughout
the paper $C_{i}~(i=1,2,3,4)$ will be reserved symbols referring to these special subcodes.

According to Definition 3.1 and [13], we have the following theorem.

\textbf{Theorem 3.1} Let $C=\eta_{1}C_{1}\oplus\eta_{2}C_{2}\oplus\eta_{3}C_{3}\oplus\eta_{4}C_{4}$ be a linear code of length $n$ over $R$. Then $C^{\bot}=\eta_{1}C_{1}^{\bot}\oplus\eta_{2}C_{2}^{\bot}\oplus\eta_{3}
C_{3}^{\bot}\oplus\eta_{4}C_{4}^{\bot}.$

According to the definition of the Gray map $\Phi$, we can easily obtain the following theorem.

\textbf{Theorem 3.2} Let $C$ be a linear code of length $n$ over $R$, $|C|=q^{k}$ and $d_{L}(C)=d$. Then $\Phi(C)$ is a $q$-ary linear code with parameter $[4n, k, d].$

Let $C=\eta_{1}C_{1}\oplus\eta_{2}C_{2}\oplus\eta_{3}C_{3}\oplus\eta_{4}C_{4}$ be a linear code of length $n$ over $R$. Since $C$ is a $F_{q}$-module, then we have the following lemma.

\textbf{Lemma 3.1} If $G_{i}$ are generator matrices of $q$-ary linear codes
$C_{i}~(i=1, 2, 3, 4)$, respectively, then the generator matrix of $C$ is
\begin{eqnarray*}
 G= \left(
      \begin{array}{cc}
   \eta_{1} G_{1} &\\
   \eta_{2} G_{2}  &\\
   \eta_{3} G_{3} &\\
     \eta_{4} G_{4} &\\
      \end{array}
    \right).
\end{eqnarray*}
Moreover, if $G_{1}=G_{2}=G_{3}$, then $G=G_{1}.$\

In light of the definition of Gray map $\Phi$, we can easily obtain the following proposition.

\textbf{Proposition 3.1} If $C$ is a linear code of length $n$ over $R$ with generator matrice $G$, then we have
\begin{equation}
\label{generator-C1} \Phi(G)=\left(
\begin{array}{cccccc}
\Phi(\eta_{1}G_{1}) \\
\Phi(\eta_{2}G_{2}) \\
\Phi(\eta_{3}G_{3}) \\
\Phi(\eta_{4}G_{4})
\end{array} \right)=\left(\begin{array}{cccccc}
G_{1} & G_{1} & G_{1} & G_{1}\\
\textbf{0} & G_{2} & \textbf{0} & G_{2}\\
\textbf{0} & \textbf{0} & G_{3} & G_{3}\\
\textbf{0} & \textbf{0} & \textbf{0} & G_{4}
\end{array}\right).\nonumber
\end{equation}

\section{Skew Cyclic codes over $\F_q+u\F_q+v\F_q+uv\F_q$ }

\hspace*{0.6cm}In this section, we assume $C_{3}$ and $C_{4}$ are equivalent. Before studying skew cyclic codes over $R$, we define a skew polynomial ring $R[X; \theta]$ and skew cyclic codes over $R$. Next, we determine the structural properties of skew cyclic codes over $R$ through a decomposition theorem.

\textbf{Definition 4.1} We define the skew polynomial ring as $R[x;\theta]=\{a_{0}+a_{1}x+\cdots+a_{n}x^{n}| a_{i}\in R,i=0,1,\cdots,n\}$, where the coefficients are written on the left of the variable $x$. The multiplication is defined by the basic rule $(ax^{i})(bx^{j})=a\theta^{i}(b)x^{i+j}$, and the
addition is defined to be the usual addition rule
of polynomials.

It is easily checked that the ring $R[x;\theta]$ is not commutative unless $\theta$ is the identity automorphism on $R$.

\textbf{Definition 4.2} A nonempty subset $C$ of $R^{n}$ is called a skew cyclic code of length $n$ if $C$ satisfies the
following conditions:
(1) $C$ is a submodule of $R^{n}$;
(2) if $r=(r_{0},r_{1},\cdots,r_{n-1})\in C,$ then skew cyclic shift $\rho(r)=(\theta(r_{n-1}),\theta(r_{0}),\cdots,\theta(r_{n-2}))\in C.$

\textbf{Theorem 4.1} Let $C=\eta_{1}C_{1}\oplus\eta_{2}C_{2}\oplus\eta_{3}C_{3}\oplus\eta_{4}C_{4}$ be a linear code of length $n$ over $R$, where $C_{i}~(i=1, 2, 3, 4)$ are codes over $\F_q$ of length $n$. Then $C$ is a skew cyclic code with respect to the automorphism $\theta$ if and only if $C_{i}$ are skew cyclic codes over $\F_q$ with respect to the automorphism $\theta$.

\textbf{Proof} For any $r=(r_{0},r_{1},\cdots,r_{n-1})\in C$, let $r_{i}=\eta_{1}a_{i}+\eta_{2}b_{i}+\eta_{3}c_{i}+\eta_{4}d_{i}$ for $0\leq i\leq n-1$, where $a=(a_{0},a_{1},\cdots,a_{n-1})\in C_{1},$ $b=(b_{0},b_{1},\cdots,b_{n-1})\in C_{2},$ $c=(c_{0},c_{1},\cdots,c_{n-1})\in C_{3}$ and $d=(d_{0},d_{1},\cdots,d_{n-1})\in C_{4}.$
If $C_{i}$ are skew cyclic codes, then $\rho(r)=\rho(\eta_{1}a+\eta_{2}b+\eta_{3}c+\eta_{4}d)=\eta_{1}\rho(a)+\eta_{2}\rho(b)+\eta_{3}\rho(c)+\eta_{4}\rho(d)\in C$. This implies that $C$ is a skew cyclic code over $R$.

On the other hand, if $C$ is a skew cyclic code over $R$, we have $\rho(r)=(\theta(r_{n-1}),\theta(r_{0}),\cdots,$ $\theta(r_{n-2}))=\eta_{1}\rho(a)
+\eta_{2}\rho(b)+\eta_{3}\rho(c)+\eta_{4}\rho(d)\in C,$ which implies $\rho(a)\in C_{1},$ $\rho(b)\in C_{2},$ $\rho(c)\in C_{3},$ $\rho(d)\in C_{4}.$ Thus $C_{i}$ are skew cyclic codes over $\F_q$.

According to [4, Corollary 18], we know that the dual code of every skew cyclic code over $\F_q$ is also skew cyclic. By using this connection and Theorem 4.1, we get the following corollary.

\textbf{Corollary 4.1} If $C$ is a skew cyclic code over $R$, then the dual code $C^{\bot}$ is also skew cyclic.

The following theorem determines the generator polynomials of a skew cyclic code of length $n$ over $R$.

\textbf{Theorem 4.2} Let $C=\eta_{1}C_{1}\oplus\eta_{2}C_{2}\oplus\eta_{3}C_{3}\oplus\eta_{4}C_{4}$ be a skew cyclic code of length $n$ over $R$ and suppose that $g_{i}(x)$ are generator polynomials of $C_{i}$~(i=1, 2, 3, 4)
respectively. Then $C=\langle\eta_{1}g_{1}(x),\eta_{2}g_{2}(x),\eta_{3}g_{3}(x),\eta_{4}g_{4}(x)\rangle$ and $|C|=q^{4n-\sum_{i=1}^{4}deg(g_{i}(x))}$.

\textbf{Proof} Since $C_{i}=\langle g_{i}(x)\rangle$, for $i=1, 2, 3, 4,$ and $C=\eta_{1}C_{1}\oplus\eta_{2}C_{2}\oplus\eta_{3}C_{3}\oplus\eta_{4}C_{4}$, then $$C=\bigg\{c(x)=\sum_{i=1}^{4}\eta_{i}r_{i}(x)g_{i}(x) | r_{i}(x)\in\F_q[x;\theta]\bigg\}.$$
Hence $C\subseteq\langle\eta_{1}g_{1}(x),\eta_{2}g_{2}(x),\eta_{3}g_{3}(x),\eta_{4}g_{4}(x)\rangle$. Conversely, for any $\sum_{i=1}^{4}\eta_{i}k_{i}(x)g_{i}(x)
\in\langle\eta_{1}g_{1}(x),\eta_{2}g_{2}(x),\eta_{3}g_{3}(x),\eta_{4}g_{4}(x)\rangle$, where $k_{i}(x)\in R[x;\theta]/(x^{n}-1)$, then there exist $ r_{i}\in\F_q[x;\theta]$ such that $\eta_{i}k_{i}(x)=\eta_{i}r_{i}(x)$, $i=1,2,3,4.$ Thus $\langle\eta_{1}g_{1}(x),\eta_{2}g_{2}(x),\eta_{3}g_{3}(x),\eta_{4}g_{4}(x)\rangle\subseteq C$, which implies $C=\langle\eta_{1}g_{1}(x),\eta_{2} g_{2}(x),\eta_{3}g_{3}(x), \eta_{4}g_{4}(x)\rangle.$ Since $|C|=|C_{1}|\cdot|C_{2}|\cdot|C_{3}|\cdot|C_{4}|$, we obtain that $|C|=q^{4n-\sum_{i=1}^{4}deg(g_{i}(x))}.$

\textbf{Theorem 4.3} Let $C_{i}~(i=1,2,3,4)$ be skew cyclic codes over $\F_q$ and $g_{i}(x)$ be the monic generator polynomials of these codes respectively, then there is a unique polynomial $g(x)\in R[x;\theta]$ such that $C=\langle g(x)\rangle$ and
$g(x)$ is a right divisor of $x^{n}-1$, where $g(x)=\sum_{i=1}^{4}\eta_{i} g_{i}(x).$

\textbf{Proof} By Theorem 4.2, we know $C=\langle\eta_{1}g_{1}(x),\eta_{2}g_{2}(x), \eta_{3}g_{3}(x),\eta_{4}g_{4}(x)\rangle.$ We take $g(x)=\eta_{1}g_{1}(x)+\eta_{2}g_{2}(x)+\eta_{3}g_{3}(x)+\eta_{4}g_{4}(x),$ obviously, we have $\langle g(x)\rangle\subseteq C$. On the other hand, one can check that $\eta_{i}g_{i}(x)=\eta_{i}g(x)(i=1,2,3,4),$ which implies $C\subseteq\langle g(x)\rangle$. Hence $C=\langle g(x)\rangle$. Since $g_{i}(x)$ are monic right divisors of $x^{n}-1 \in \F_q[x;\theta]$, then there exist $r_{i}(x)\in \F_q[x;\theta]$ such that $x^{n}-1=r_{i}(x)g_{i}(x)$. Thus
\begin{eqnarray*}
[\eta_{1}r_{1}(x)+\eta_{2}r_{2}(x)+\eta_{3}r_{3}(x)+\eta_{4}r_{4}(x)]g(x)&=& \sum_{i=1}^{4}\eta_{i}r_{i}(x)\cdot
\sum_{i=1}^{4}\eta_{i}g_{i}(x) \\
   &=&\sum_{i=1}^{4} \eta_{i}r_{i}(x)g_{i}(x) \\
   &=& \sum_{i=1}^{4}\eta_{i}(x^{n}-1) \\
   &=& x^{n}-1.
\end{eqnarray*}
This implies $g(x)$ is a right divisor of $x^{n}-1$.

\textbf{Corollary 4.2} Every left submodule of $R[x;\theta]/(x^{n}-1)$ is principally generated.

Let $g(x)=g_{0}+g_{1}x+\cdots+g_{t}x^{t}$ and $h(x)=h_{0}+h_{1}x+\cdots+h_{n-t}x^{n-t}$ be polynomials in $\F_q[x;\theta]$ such that $x^{n}-1=h(x)g(x)$ and $C$ be the skew cyclic code generated by $g(x)$ in
$\F_q[x;\theta]/(x^{n}-1)$, according to Corollary 18 in [4], then the dual code of $C$ is a skew cyclic code generated by $\widetilde{h}(x)=h_{n-t}+\theta(h_{n-t-1})x+\cdots+\theta^{n-t}(h_{0})x^{n-t}$. Therefore we have the following corollary.

\textbf{Corollary 4.3} Let $C_{i}$ be skew cyclic codes over $\F_q$ and $g_{i}(x)$ be their generator polynomial such that $x^{n}-1=h_{i}(x)g_{i}(x)$ in $\F_q[x;\theta]$. If $C$ is a skew cyclic code over $R$, then $C^{\bot}=\langle\sum_{i=1}^{4}\eta_{i}\widetilde{h_{i}}(x)\rangle$ and $|C^{\bot}|=q^{\sum_{i=1}^{4}deg(g_{i}(x))}$.

In light of previous introduction, we know that the order of $\theta$ is even. Therefore, we always assume that $n$ be odd in the rest of the paper.

\textbf{Theorem 4.4} [6] Let $n$ be odd and $C$ be a skew cyclic code of length $n$, then $C$ is equivalent to a cyclic code of length $n$ over $R$.

By Theorem 4.4, we can determine the number of distinct skew cyclic codes of odd length $n$ over $R$.

\textbf{Corollary 4.4} Let $n$ be odd and $x^{n}-1=\prod_{i=1}^{r}p_{i}^{s_{i}}(x)$, where $p_{i}(x)\in F_{q}[x;\theta_{i}]$ is irreducible, then the number of distinct skew cyclic codes of length $n$ over $R$ is equal to the number of ideals in $R[x]/(x^{n}-1)$, i.e. $\prod_{i=1}^{r}(s_{i}+1)^{4}$.

\section{Application Examples}
\hspace*{0.6cm}In this section, we will exhibit a example of skew cyclic codes and their Gray images over $GF(9)$. Before giving a example, we first give the definition of Plotkin Sum.

Let $C\oplus_P D$ denote the Plotkin sum of two linear codes $C$ and $D$, also called $(u|u + v)$ construction, where $u\in C, v\in D$.
For more information on the Plotkin sum, one can see a good survey [9].

In the following, we assume $G_{i}$ are generator matrices of $9$-ary linear codes $C_{i}$ for $i=1,2,3,4,$ respectively. Let $C=\eta_{1}C_{1}\oplus\eta_{2}C_{2}\oplus\eta_{3}C_{3}\oplus\eta_{4}C_{4}$ be a linear code of length $n$ over $R$, then its Gray image $\Phi(C)$ is none other than $$(C_1\oplus_P C_2)\oplus_P(C_3\oplus_P C_4).$$
We construct skew cyclic codes over $GF(9)$ with some conditions. If $C_1$ is a $[20,1,20]$ code, $C_2$ is a $[20,9,4]$ code, $C_3$ is a $[20,10,2]$ code and $C_4$ is a $[20,10,2]$ code, then the Gray image of $C$ has parameters $[80,30,4]$ over $GF(9).$

\section{Conclusion}

\hspace*{0.6cm}This paper is devoted to studying skew cyclic codes over $R=\F_q+u\F_q+v\F_q+uv\F_q,$ where $u^{2}=u,v^{2}=v,uv=vu, q=p^{m}$ and $p$ is an odd prime.
 First, we introduce the structure of linear codes over $R$ and  show the structural properties of skew cyclic codes over $R$. Next, we give the enumeration of distinct skew cyclic codes over $R$ when $n$ is odd.

\end{CJK}

\end{document}